\documentclass[12pt]{article}

\usepackage{amsmath}
\usepackage{amssymb}
\usepackage{indentfirst}
\usepackage[dvips,xdvi]{graphicx}

\newtheorem{theorem}{Theorem}
\newtheorem{lemma}{Lemma}

\newtheorem{proposition}{Proposition}

\newcommand{\R}{\ensuremath{{\mathbb R}}} 
\renewcommand{\epsilon}{\varepsilon}
\renewcommand{\phi}{\varphi}

\newcommand{\Def}{\ensuremath{\stackrel{\rm def}=}} 

\newcommand{\ST}{\ensuremath{\quad\text{such that}\quad}}
\newcommand{\AND}{\ensuremath{\quad\text{and}\quad}}

\renewcommand{\H}{\ensuremath{{\cal H}}}
\renewcommand{\S}{\ensuremath{{\cal S}}}

\renewcommand{\L}{\ensuremath{{\cal L}}}
\newcommand{\X}{\ensuremath{{\cal X}}}

\newcommand{\E}{\ensuremath{{\cal E}}}

\newcommand{\Hn}{\ensuremath{{\cal H}^{\otimes n}}}
\newcommand{\rhon}{\ensuremath{\rho^{\otimes n}}}
\newcommand{\sigman}{\ensuremath{\sigma^{\otimes n}}}

\newcommand{\psibar}{\ensuremath{\overline{\psi}}}
\newcommand{\phibar}{\ensuremath{\overline{\phi}}}

\newcommand{\Tr}{\ensuremath{\mbox{\rm Tr}}}

\newcommand{\argmax}{\ensuremath{\mathop{\rm arg \; max}}}

\newcommand{\proj}[1]{\ensuremath{\left\{ #1 \right\}}}
\newcommand{\oneover}[1]{\ensuremath{\frac{1}{#1}}}
\newcommand{\defset}[2]{\ensuremath{%
 \left\{#1\,\left|\,#2\right.\right\}
}}

\newcommand{\pinching}[2]{\ensuremath{\,\E_{#1}\!\!\left(#2\right)}}

\newcommand{\bigzeroUR}{{\lower0.8ex\hbox{\Large 0}}}
\newcommand{\bigzeroDL}{{\raise0.8ex\hbox{\Large 0}}}

\def\QED{\mbox{\rule[0pt]{1.5ex}{1.5ex}}}

\def\endproof{\hspace*{\fill}~\QED\par\endtrivlist\unskip}
\def\keywords{\vspace{-.3em}
    \if@twocolumn
      \small\it Keywords\/\bf---$\!$%
    \else
      \begin{center}\small\bf Keywords\end{center}\quotation\small
    \fi}
\def\endkeywords{\vspace{0.6em}\par\if@twocolumn\else\endquotation\fi
    \normalsize\rm}
\def\appendix{\par
    \setcounter{section}{0}\setcounter{subsection}{0}
    \def\thesection{\Alph{section}} \section*{Appendix}
}
\def\appendices{\par
    \setcounter{section}{0}\setcounter{subsection}{0}
    \def\thesection{\Alph{section}} \section*{Appendices}
}

\begin{document}

\bibliographystyle{IEEE}

\title{
On Error Exponents in Quantum Hypothesis Testing
}

\author{
Tomohiro Ogawa
\thanks{
Department of Mathematical Informatics,
Graduate School of Information Science and Technology, University of Tokyo,
7--3--1 Hongo, Bunkyo-ku, Tokyo, 113--8656, Japan.
(e-mail: ogawa@sr3.t.u-tokyo.ac.jp)
},
\and
Masahito Hayashi
\thanks{
Laboratory for Mathematical Neuroscience,
Brain Science Institute, RIKEN,
2--1 Hirosawa, Wako, Saitama, 351--0198, Japan.
(e-mail: masahito@brain.riken.go.jp)
}
}

\date{}

\maketitle

\begin{abstract}
In the simple quantum hypothesis testing problem,
upper bounds on the error probabilities are shown
based on a key operator inequality
between a density operator and its pinching.
Concerning the error exponents,
the upper bounds lead to
a noncommutative analogue of the Hoeffding bound,
which is identical with the classical counter part
if the hypotheses,
composed of two density operators, are mutually commutative.
The upper bounds also provide
a simple proof of the direct part of the quantum Stein's lemma.
\end{abstract}


\section{Introduction}
\label{sec:intro}

Quantum hypothesis testing is a fundamental problem
in quantum information theory,
because it is one of the most simple problems
where the difficulty derived from
noncommutativity of operators appears.
It is also closely related to other topics in quantum
information theory, as in classical information theory.
Actually, its relation with quantum channel coding is discussed
in \cite{Ogawa-Nagaoka-ISIT2002} \cite{Hayashi-Nagaoka-ISIT2002}.

Let us outline briefly significant results in
classical hypothesis testing
for probability distributions $p^n(\cdot)$ versus $q^n(\cdot)$,
where $p^n(\cdot)$ and $q^n(\cdot)$ are
independently and identically distributed (i.i.d.) extensions
of some probability distributions $p(\cdot)$ and $q(\cdot)$
on a finite set $\X$.
In the classical case,
the asymptotic behaviors of the first kind error probability
$\alpha_n$ and the second kind error probability $\beta_n$
for the optimal test were studied thoroughly as follows.

First, when $\alpha_n$
satisfies the constant constraint $\alpha_n\le\epsilon$ $(\epsilon>0)$,
the error exponent of $\beta_n$ for the optimal test is written
asymptotically as
\begin{align}
\lim_{n\rightarrow\infty}\oneover{n}\log\beta_n = -D(p||q)
\label{classical-Stein}
\end{align}
for any $\epsilon$, where $D(p||q)$ is the Kullback-Leibler divergence.
The equality \eqref{classical-Stein} is called
Stein's lemma (see {\it e.g.} \cite{Blahut-text}, p.115),
and the quantum analogue of \eqref{classical-Stein} was established
recently \cite{Hiai-Petz} \cite{Ogawa-Nagaoka-2000}.

Next, when $\alpha_n$ satisfies the exponential constraint
$\alpha_n\le e^{-nr}$ $(r> 0)$, the error exponent of $\beta_n$
for the optimal test is asymptotically determined by
\begin{align}
\limsup_{n\rightarrow\infty}\oneover{n}\log\beta_n
&= -\min_{p':\,D(p'||p)\le r} D(p'||q)
\label{classical-Hoeffding-1}
\\
&= -\max_{0 < s \le 1}\frac{\Psi(s)-(1-s)r}{s},
\label{classical-Hoeffding}
\end{align}
where the function $\Psi(s)$ is defined as
\begin{align}
\Psi(s) \Def \sum_{x\in\X} p(x)^{1-s} q(x)^{s}.
\label{classical-psi}
\end{align}
Historically speaking,
\eqref{classical-Hoeffding-1} and the test achieving it were shown
in \cite{Hoeffding}, followed by another expression
\eqref{classical-Hoeffding} (see \cite{Blahut}),
which we call the Hoeffding bound here.
In quantum hypothesis testing,
the error exponent of $1-\beta_n$ was studied in \cite{Ogawa-Nagaoka-2000}
to obtain a similar result to \eqref{classical-Hoeffding}, 
which led to the strong converse property in quantum hypothesis testing.
Concerning quantum fixed-length pure state source coding,
the error exponet of erroneously decoded probability
was determined in \cite{Hayashi-fixed-length-source},
where the optimality of the error exponent
similar to \eqref{classical-Hoeffding}
was discussed.

In this manuscript, a quantum analogue of the Hoeffding bound
\eqref{classical-Hoeffding} \eqref{classical-psi} is introduced
to derive a bound on the error exponent in quantum hypothesis testing.
As a by-product of the process to derive the exponent,
a simple proof of the quantum Stein's lemma is also given.

\section{Definition and Main Results}
\label{sec:definition}

Let $\H$ be a Hilbert space which represents a physical system in interest.
We assume $\dim\H<\infty$ for mathematical simplicity.
Let us denote the set of linear operators on $\H$ as $\L(\H)$
and define the set of density operators on $\H$ by 
\begin{align}
\S(\H) \Def \defset{\rho\in\L(\H)}{\rho=\rho^*\ge 0, \Tr[\rho]=1}.
\end{align}
We study the hypothesis testing problem for
the null hypothesis $H_0 : \rho_n\Def\rhon\in\S(\Hn)$
versus the alternative hypothesis $H_1 : \sigma_n\Def\sigman\in\S(\Hn)$,
where $\rhon$ and $\sigman$ are the $n$th
tensor powers of arbitrarily given density operators 
$\rho$ and $\sigma$ in $\S(\H)$.

The problem is to decide which hypothesis is true
based on the data drawn from a quantum measurement,
which is described by a positive operator valued measure (POVM)
on $\Hn$, i.e.
a resolution of identity $\sum_i M_{n,i} = I_n$
by nonnegative operators $M_n=\{M_{n,i}\}$ on $\Hn$.
If a POVM consists of projections on $\Hn$,
it is called a projection valued measure (PVM).
In the hypothesis testing problem, however,
it is sufficient to treat a two-valued POVM $\{M_0,M_1\}$,
where the subscripts $0$ and $1$ indicate the acceptance
of $H_0$ and $H_1$, respectively.
Thus, an operator $A_n\in\L(\Hn)$ satisfying inequalities
$0\le A_n\le I_n$ is called a test
in the sequel, since $A_n$ is 
identified with the POVM $\{A_n,\, I_n-A_n\}$. 
For a test $A_n$, the error probabilities of the first kind and 
the second kind are, respectively, defined by
\begin{align*}
\alpha_n (A_n) &\Def \Tr[\rho_n (I_n -A_n)],
\\
\beta_n (A_n) &\Def \Tr[\sigma_n A_n].
\end{align*}

Let us define the optimal value for $\beta_n(A_n)$
under the constant constraint on $\alpha_n(A_n)$:
\begin{align}
\beta_n^*(\epsilon)\Def\min
\bigl\{ \beta_n(A_n) &\bigm| A_n:\text{test},\,
\alpha_n(A_n)\le\epsilon \bigr\},
\end{align}
and let
\begin{align}
D(\rho\|\sigma)\Def\Tr[\rho(\log\rho-\log\sigma)],
\end{align}
which is called the quantum relative entropy.
Then we have the following theorem, which is one of the most
essential theorems in quantum information theory.
\vspace{2ex}
\begin{proposition}[The quantum Stein's lemma]
For $0<\forall\epsilon<1$, it holds that
\begin{align}
\lim_{n\rightarrow\infty}\oneover{n}\log\beta_n^*(\epsilon)=-D(\rho\|\sigma).
\label{Stein}
\end{align}
\end{proposition}
\vspace{2ex}
The first proof of \eqref{Stein} was composed of two inequalities,
the direct part and the converse part.
The direct part,
concerned with existence of good tests,
claims that
\begin{align}
0<\forall\epsilon\le 1,\quad
\limsup_{n\rightarrow\infty}\oneover{n}\log\beta_n^*(\epsilon)
\le -D(\rho\|\sigma),
\label{direct1}
\end{align}
and it was given by Hiai-Petz \cite{Hiai-Petz}.
In this manuscript, the main focus is on the direct part,
which is sometimes referred to
as an equivalent form (see \cite{Ogawa-Nagaoka-2000}):
\begin{align}
&\exists \{A_n:\text{test}\}_{n=1}^{\infty}
\ST
\lim_{n\rightarrow\infty}\alpha_n(A_n)=0
\nonumber
\\
&\AND
\limsup_{n\rightarrow\infty}\oneover{n}\log\beta_n(A_n)
\le -D(\rho\|\sigma).
\label{direct2}
\end{align}
On the other hand, the converse part,
concerned with nonexistence of too good tests,
asserts that
\begin{align}
0\le\forall\epsilon<1,\quad
\liminf_{n\rightarrow\infty}\oneover{n}\log\beta_n^*(\epsilon)
\ge -D(\rho\|\sigma),
\label{converse1}
\end{align}
which was given by Ogawa-Nagaoka \cite{Ogawa-Nagaoka-2000}.
A direct proof of the equality \eqref{Stein}
was also given by Hayashi \cite{Hayashi-Stein}
using the information spectrum approach in quantum setting \cite{Nagaoka},
and a considerably simple proof of the converse part \eqref{converse1}
was given in \cite{Nagaoka-converse}, recently.

In this manuscript, the asymptotic behavior of the error exponent
$\oneover{n}\log\beta_n(A_n)$ under the exponential constraint
$\alpha_n(A_n)\le e^{-nr}$ $(r>0)$ is studied,
and a noncommutative analogue of the Hoeffding bound \cite{Hoeffding}
similar to \eqref{classical-Hoeffding} is given as follows.
\vspace{2ex}
\begin{theorem}
For $\forall r>0$, there exists a test $A_n$ which satisfies
\begin{align}
\limsup_{n\to\infty}\oneover{n}
\log\alpha_n\left(A_n\right)
&\le -r,
\label{theorem-rate-1} \\
\limsup_{n\to\infty}\oneover{n}
\log\beta_n\left(A_n\right)
&\le -\max_{0 < s \le 1}\frac{\psibar(s)-(1-s)r}{s},
\label{theorem-rate-2}
\end{align}
where
\begin{align}
\psibar(s)&\Def -\log\Tr\left[
\rho\,\sigma^{\frac{s}{2}}\rho^{-s}\sigma^{\frac{s}{2}}\right].
\label{def-psibar}
\end{align}
\label{theorem:rate}
\end{theorem}
\vspace{-2ex}
We will prove the theorem in Section \ref{sec:exponent}.
If $\rho$ and $\sigma$ are mutually commutative,
$\psibar(s)$ is identical with the classical counterpart
$\Psi(s)$ defined in \eqref{classical-psi},
and \eqref{theorem-rate-2}
coincides with the Hoeffding bound \eqref{classical-Hoeffding},
which is optimal in classical hypothesis testing.

This manuscript is organized as follows.
In Section \ref{sec:bounds},
upper bounds on the error probabilities
are shown based on a key operator inequality \cite{Hayashi-Stein}.
Using the upper bounds,
we will prove Theorem \ref{theorem:rate} in Section \ref{sec:exponent}.
In Section \ref{sec:graph},
the behavior of the function \eqref{def-psibar} is investigated,
and a simple proof of
the direct part of the quantum Stein's lemma \eqref{direct2}
is given as a consequence of the upper bounds.

Two appendices are included for readers' convenience.
Appendix~\ref{appendix:pinching} is devoted to the definition
of the pinching map used effectively in Section \ref{sec:bounds}.
In Appendix~\ref{appendix:key-inequality},
the key operator inequality used in Section \ref{sec:bounds}
is summarized briefly,
along with another proof of it for readers' convenience.

\section{Bounds on Error Probabilities}
\label{sec:bounds}

In the sequel,
let $\pinching{\sigma_n}{\rho_n}$ be the pinching 
defined in Appendix~\ref{appendix:pinching}
and denote it
as $\overline{\rho_n}$ for simplicity.
Let $v(\sigma_n)$ be the number of eigenvalues
of $\sigma_n$ mutually different from others
as defined in Appendix~\ref{appendix:pinching}.
Then a key operator inequality
\footnote{
Although the way to derive the operator inequality
and the definition of $v(\sigma_n)$
are different from those of \cite{Hayashi-Stein},
it results in the same one as \cite{Hayashi-Stein}
in the case that both of $\rho_n$ and $\sigma_n$
are tensored states.
}
follows from Lemma~\ref{lemma:Hayashi}
in Appendix~\ref{appendix:key-inequality},
which was originally appeared in \cite{Hayashi-Stein}:
\begin{align}
\rho_n \le v(\sigma_n)\,\overline{\rho_n}.
\label{key-inequality}
\end{align}
Note that the type counting lemma
(see {\it e.g.} \cite{Cover}, Theorem 12.1.1) provides
\begin{align}
v(\sigma_n)\le (n+1)^d,
\label{type-counting}
\end{align}
where $d\Def\dim\H$.
Following \cite{Hayashi-Stein},
let us apply the operator monotonicity of
the function $x \longmapsto -x^{-s}$ $(0\le s\le 1)$
(see {\it e.g} \cite{Bhatia})
to the key operator inequality \eqref{key-inequality}
so that we have
\begin{align}
\overline{\rho_n}^{\,-s}
&\le v(\sigma_n)^{s} \rho_n^{-s}
\nonumber \\
&\le (n+1)^{sd} \rho_n^{-s} .
\label{h1}
\end{align}
Here, let us define
the projection $\proj{ X > 0 }$
for a Hermitian operator $X= \sum_i x_i E_i$ as
\begin{align}
\proj{ X > 0 } \Def \sum_{i:x_i > 0} E_i,
\end{align}
where $E_i$ is the projection onto the eigenspace
corresponding to an eigenvalue $x_i$.
With the above notation, we will focus on a test defined as
\begin{align}
\overline{S}_n(a) \Def \proj{ \overline{\rho_n} - e^{na}\sigma_n > 0 },
\label{def-test}
\end{align}
where $a$ is a real parameter,
and derive the upper bounds on the error probabilities
for the test $\overline{S}_n(a)$ as follows.
\vspace{2ex}
\begin{theorem}
\begin{align}
\alpha_n\left(\overline{S}_n(a)\right)
&\le (n+1)^{d}\, e^{-n\phibar(a)},
\label{theorem-alpha} \\
\beta_n\left(\overline{S}_n(a)\right)
&\le (n+1)^{d}\, e^{-n\left[\phibar(a)+a\right]},
\label{theorem-beta}
\end{align}
where $\phibar(a)$ is defined by
$\psibar(s)$ given in \eqref{def-psibar} as
\begin{align}
\phibar(a)&\Def\max_{0\le s\le 1}\left\{ \psibar(s) - as \right\}.
\label{def-phibar}
\end{align}
\label{theorem-bound}
\end{theorem}
\vspace{-1ex}
\begin{proof}
The definition of $\overline{S}_n(a)$ and commutativity of operators
$\overline{\rho_n}$ and $\sigma_n$ lead to
\begin{align}
&\left(\overline{\rho_n}^{1-s} - e^{na(1-s)}\sigma_n^{1-s}\right)
\overline{S}_n(a) \ge 0,
\label{stein-new-1}
\\
&\left(\overline{\rho_n}^s - e^{nas}\sigma_n^s\right)
\left(I_n-\overline{S}_n(a)\right)\le 0
\label{stein-new-2}
\end{align}
for $\forall s\ge 0$.
Note that $\overline{S}_n(a)$ also commutes with $\sigma_n$.
Therefore, the inequality \eqref{stein-new-2},
with the property of the pinching \eqref{pinching2}
in Appendix~\ref{appendix:pinching}, provides
\begin{align}
\alpha_n\left(\overline{S}_n(a)\right)
&=\Tr\left[\rho_n\left(I_n-\overline{S}_n(a)\right)\right]
\nonumber\\
&=\Tr\left[\overline{\rho_n}\left(I_n-\overline{S}_n(a)\right)\right]
\nonumber\\
&= \Tr\left[\overline{\rho_n}^{\,1-s} \overline{\rho_n}^{\,s}
\left(I_n-\overline{S}_n(a)\right)\right]
\nonumber\\
&\le e^{nas}\Tr\left[\overline{\rho_n}^{\,1-s}\sigma_n^s
\left(I_n-\overline{S}_n(a)\right)\right]
\nonumber\\
&\le e^{nas}\Tr\left[\overline{\rho_n}^{\,1-s}\sigma_n^s\right].
\label{stein-new-3}
\end{align}
In the same way, \eqref{stein-new-1} yields
\begin{align}
\beta_n\left(\overline{S}_n(a)\right)
&=\Tr\left[\sigma_n\overline{S}_n(a)\right]
\nonumber\\
&=\Tr\left[\sigma_n^{s}\sigma_n^{1-s}\overline{S}_n(a)\right]
\nonumber\\
&\le e^{-na(1-s)}\Tr\left[\sigma_n^{s}
\overline{\rho_n}^{\,1-s}\overline{S}_n(a)\right]
\nonumber\\
&\le e^{-na}e^{nas}\Tr\left[\overline{\rho_n}^{\,1-s}\sigma_n^s\right].
\label{stein-new-4}
\end{align}
Here, it follows from the property \eqref{pinching2} and \eqref{h1} that
\begin{align}
\Tr\left[\overline{\rho_n}^{\,1-s}\sigma_n^s\right]
&=\Tr\left[\overline{\rho_n}\,\sigma_n^{\frac{s}{2}}\,
\overline{\rho_n}^{\,-s}\sigma_n^{\frac{s}{2}}\right]
\nonumber\\
&=\Tr\left[\rho_n\sigma_n^{\frac{s}{2}}\,\overline{\rho_n}^{\,-s}
\sigma_n^{\frac{s}{2}}\right]
\nonumber\\
&\le (n+1)^{sd}\,
\Tr\left[\rho_n\sigma_n^{\frac{s}{2}}\,\rho_n^{\,-s}
\sigma_n^{\frac{s}{2}}\right]
\nonumber\\
&= (n+1)^{sd}\,
\left(
\Tr\left[\rho\,\sigma^{\frac{s}{2}}\rho^{-s}\sigma^{\frac{s}{2}}\right]
\right)^n
\nonumber\\
&\le (n+1)^{d}\, e^{-n\psibar(s)}
\label{stein-new-5}
\end{align}
for $0\le \forall s\le 1$.
Combining \eqref{stein-new-3} \eqref{stein-new-4} \eqref{stein-new-5},
we have
\begin{align}
\alpha_n\left(\overline{S}_n(a)\right)
&\le (n+1)^{d}\, e^{-n\left[ \psibar(s) - as \right]},
\\
\beta_n\left(\overline{S}_n(a)\right)
&\le (n+1)^{d}\, e^{-n\left[ \psibar(s) - as +a \right]},
\end{align}
which lead to \eqref{theorem-alpha} \eqref{theorem-beta}
by taking the maximum in the exponents.
\end{proof}

\section{Proof of Theorem \ref{theorem:rate}}
\label{sec:exponent}

In this section, we will prove Theorem \ref{theorem:rate}
after preparing two lemmas,
where the behavior of $\phibar(a)$
in the error exponents \eqref{theorem-alpha} \eqref{theorem-beta}
is investigated.
\vspace{2ex}
\begin{lemma}
$\phibar(a)$ is convex and monotonically nonincreasing.
\label{lamma-convex-monotone}
\end{lemma}
\vspace{2ex}
\begin{proof}
The assertion immediately follows from the definition of $\phibar(a)$.
Actually, we have for $0\le\forall t\le 1$
\begin{align}
\phibar(ta+(1-t)b)
&=\max_{0\le s\le 1}\left\{\psibar(s)- (ta+(1-t)b) s \right\}
\nonumber \\
&\le t \max_{0\le s\le 1}\left\{\psibar(s)- a s \right\}
+(1-t)\max_{0\le s\le 1}\left\{\psibar(s)- b s \right\}
\nonumber \\
&= t \phibar(a) + (1-t) \phibar(b).
\end{align}
Next, let $a\le b$
and
$s_b\Def\argmax_{0\le s\le 1}\left\{\psibar(s)-bs\right\}$.
Then we have
\begin{align}
\phibar(b)
&=\psibar(s_b) -bs_b
\nonumber \\
&\le \psibar(s_b) -as_b
\nonumber \\
&\le \max_{0\le s\le 1}\left\{\psibar(s) -as\right\}
\nonumber \\
&=\phibar(a).
\end{align}
\end{proof}


\vspace{2ex}
\begin{lemma}
$\phibar(a)$ ranges from $0$ to infinity.
\label{lamma-range}
\end{lemma}
\vspace{2ex}
\begin{proof}
Since we can calculate the derivative of $\psibar(s)$ explicitly,
$\psibar(s)$ is continuous and differentiable.
Therefore, it follows from the mean value theorem that
for $s>0$ there exists $0\le t\le s$ such that
\begin{align}
\psibar'(t)=\frac{\psibar(s)-\psibar(0)}{s-0}.
\end{align}
Let $a\ge\max_{0\le t\le 1}\psibar'(t)$, then we have
\begin{align}
a\ge\frac{\psibar(s)-\psibar(0)}{s-0},
\end{align}
and hence
\begin{align}
\psibar(0)\ge\psibar(s)-as,
\end{align}
which yields
\begin{align}
0=\psibar(0)=\max_{0\le s\le 1}\left\{ \psibar(s) - as \right\}=\phibar(a).
\label{phibar-zero}
\end{align}
On the other hand, it is obvious that
\begin{align}
\lim_{a\longrightarrow-\infty}\phibar(a)=\infty.
\label{phibar-infty}
\end{align}
Since $\phibar(a)$ is continuous,
which follows from convexity by Lemma \ref{lamma-convex-monotone},
the assertion follows from \eqref{phibar-zero} and \eqref{phibar-infty}.
\end{proof}
\vspace{2ex}

Combined with the above lemma,
Theorem \ref{theorem-bound} leads to
Theorem \ref{theorem:rate} as follows.

\vspace{2ex}
\noindent{\it Proof of Theorem \ref{theorem:rate}:}\quad
For $\forall r>0$,
there exists $a_r\in\R$ such that $r=\phibar(a_r)$
from Lemma \ref{lamma-range}.
Let $\overline{u}(r)\Def\phibar(a_r)+a_r$, then
it follows from Theorem \ref{theorem-bound} that
\begin{align}
\limsup_{n\to\infty}\oneover{n}
\log\alpha_n\left(\overline{S}_n(a_r)\right)
&\le -r,
\label{Sn-ar-1)} \\
\limsup_{n\to\infty}\oneover{n}
\log\beta_n\left(\overline{S}_n(a_r)\right)
&\le -\overline{u}(r).
\label{Sn-ar-2)}
\end{align}
Therefore, it suffices to show that
\begin{align}
\overline{u}(r)
=\max_{0<s\le 1}\frac{\psibar(s)-(1-s)r}{s}.
\label{rate-proof-u}
\end{align}
For $0\le\forall s\le 1$,
we have from the definition of $\phibar(a)$
\begin{align}
r=\phibar(a_r)\ge\psibar(s)-a_rs,
\label{rate-proof-1}
\end{align}
and there exists a number $s_0$ $(0< s_0\le 1)$ achieving the equality
since $r=\phibar(a_r)>0$.
On the other hand, the definitions of $u(r)$ and $a_r$ lead to
\begin{align}
\overline{u}(r)=\phibar(a_r)+a_r=r+a_r.
\label{rate-proof-2}
\end{align}
Eliminating $a_r$ from \eqref{rate-proof-1} and \eqref{rate-proof-2},
we have
\begin{align}
\overline{u}(r)\ge\frac{\psibar(s)-(1-s)r}{s},
\label{rate-proof-3}
\end{align}
and $s_0$ achieves the equality in \eqref{rate-proof-3} as well.
Thus, we have shown \eqref{rate-proof-u},
and Theorem \ref{theorem:rate} has been proved.
\endproof

\section{Graphs of $\psibar(s)$ and $\phibar(a)$}
\label{sec:graph}

In this section,
we will investigate the graphs of $\psibar(s)$ and $\phibar(a)$.
To this end, let us define
\begin{align}
\psi(s)&\Def -\log\Tr\left[\rho^{1-s}\sigma^{s}\right],
\label{def-psi} \\
\phi(a)&\Def \max_{0\le s\le 1} \left\{ \psi(s) - as \right\}.
\label{def-phi}
\end{align}
Then we have the following lemma.
\vspace{2ex}
\begin{lemma}
\begin{align}
&\psibar(s)\le\psi(s)\quad (0\le \forall s\le 1),
\label{psi-order} \\
&\phibar(a)\le\phi(a)\quad (\forall a\in \R).
\label{phi-order}
\end{align}
\label{lamma-order}
\end{lemma}
\vspace{-4ex}
\begin{proof}
Let us apply
the monotonicity property of the quantum quasi-entropy
\cite{Petz_RIMS} \cite{Petz}
to $\Tr\left[\rho^{1-s}\sigma^{s}\right]$ $(0\le s\le 1)$
\footnote{
A comprehensible explanation of the monotonicity property is found
in \cite{Ogawa-Nagaoka-2000}.
}
so that we have
\begin{align}
e^{-n\psi(s)}
&=\left(\Tr\left[\rho^{1-s}\sigma^{s}\right]\right)^{n}
\nonumber \\
&=\Tr\left[\rho_n^{1-s}\sigma_n^{s}\right]
\nonumber \\
&\le\Tr\left[\overline{\rho_n}^{\,1-s}\sigma_n^{s}\right]
\nonumber \\
&\le (n+1)^{sd}\, e^{-n\psibar(s)},
\end{align}
where we used \eqref{stein-new-5} in the last inequality.
Thus, we obtain
\begin{align}
\psibar(s) \le \psi(s) + \frac{sd}{n}\log(n+1)
\end{align}
for any positive number $n$,
and we have \eqref{psi-order}
by letting $n$ go to infinity.
Now \eqref{phi-order} is obvious from the definition of $\phibar(a)$.
Actually, let $s_a\Def\argmax_{0\le s\le 1}\left\{\psibar(s)-as\right\}$,
then we have
\begin{align}
\phibar(a)&=\psibar(s_a)-as_a
\nonumber \\
&\le\psi(s_a)-as_a
\nonumber \\
&\le\max_{0\le s\le 1}\left\{ \psi(s)-as \right\}
\nonumber \\
&=\phi(a).
\end{align}
\end{proof}
\vspace{2ex}

\begin{figure}[htbp]
\begin{center}
\includegraphics[width=0.6\textwidth,clip]{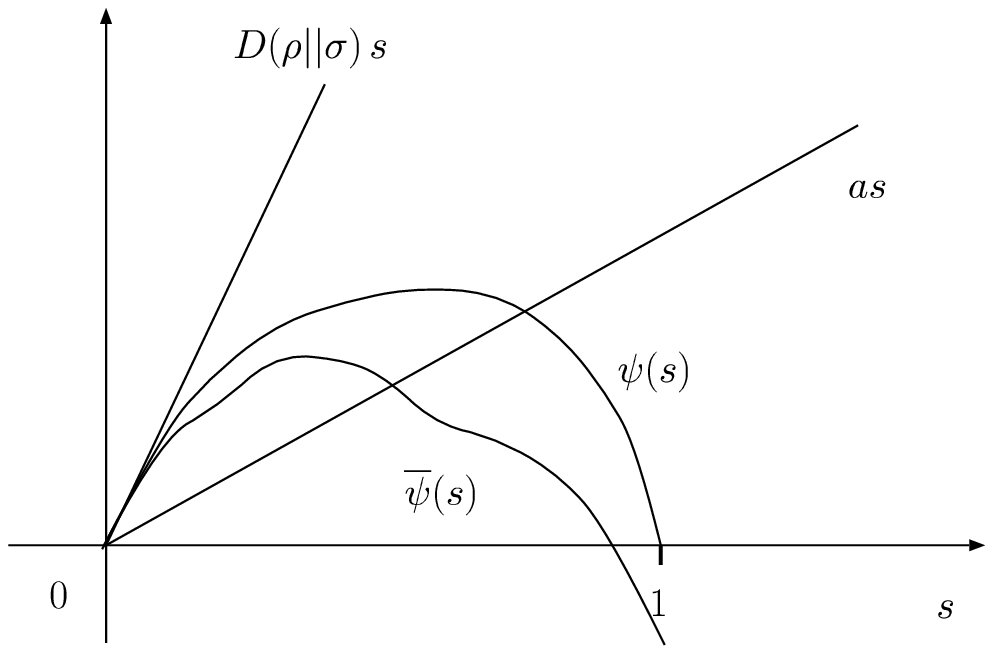}
\end{center}
\caption{The graph of $\psibar(s)$}
\label{fig:psi}
\end{figure}

\begin{figure}[htbp]
\begin{center}
\includegraphics[width=0.6\textwidth,clip]{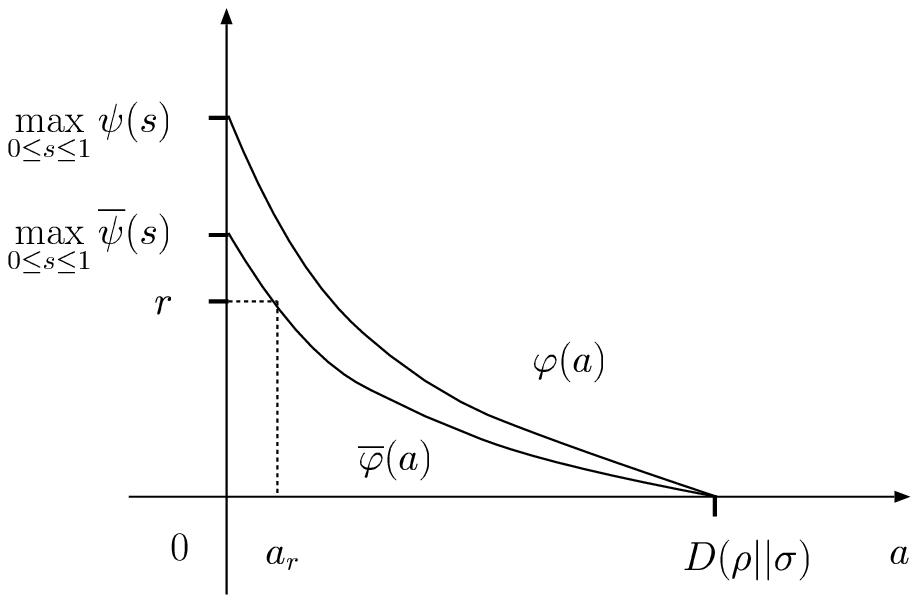}
\end{center}
\caption{The graph of $\phibar(a)$}
\label{fig:phi}
\end{figure}

Following \cite{Ogawa-Nagaoka-2000},
we can easily draw the graphs of $\psi(s)$ and $\phi(a)$
(see Figure \ref{fig:psi} and \ref{fig:phi})
by calculating the derivatives
\begin{align}
\psi'(s)&=
e^{\psi(s)}\Tr\left[\rho^{1-s}\sigma^{s}
\left(\log\rho-\log\sigma\right)\right],
\\
\psi''(s)&=-e^{\psi(s)}\Tr\left[\rho^{1-s}A\sigma^{s}A\right]
\nonumber \\
&=-e^{\psi(s)}\Tr\left[
\left( \rho^{\frac{1-s}{2}}A\sigma^{\frac{s}{2}} \right)
\left( \rho^{\frac{1-s}{2}}A\sigma^{\frac{s}{2}} \right)^*
\right]
\nonumber \\
&< 0,
\end{align}
where we put
\begin{align}
A\Def\log\rho-\log\sigma-\psi'(s).
\end{align}
Especially, note that $\psi(0)=0$ and $\psi'(0)=D(\rho||\sigma)$.

On the other hand, we can not know a lot
concerning the graphs of $\psibar(s)$ and $\phibar(a)$,
except that we have $\psibar(0)=0$ and $\psibar'(0) = D(\rho\|\sigma)$.
Considering lemmas from \ref{lamma-convex-monotone} to \ref{lamma-order},
however, we can show the graphs of $\psibar(s)$ and $\phibar(a)$ roughly
as Figure \ref{fig:psi} and \ref{fig:phi}.
Here, it should be pointed out that
\begin{align}
\phibar(a) >0 \quad \mbox{for} \quad \forall a<D(\rho||\sigma),
\end{align}
which leads to the following theorem
combined with Theorem \ref{theorem-bound}.
\vspace{2ex}
\begin{theorem}
For $\forall a<D(\rho||\sigma)$, we have
\begin{align}
&\lim_{n\rightarrow\infty}\alpha_n\left(\overline{S}_n(a)\right)=0,
\\
&\limsup_{n\rightarrow\infty}\oneover{n}
\log\beta_n\left(\overline{S}_n(a)\right) \le -a.
\end{align}
\end{theorem}
\vspace{2ex}
Since $a<D(\rho||\sigma)$ can be arbitrarily near $D(\rho||\sigma)$,
we have shown the direct part of the quantum Stein's lemma \eqref{direct2}.

\section{Concluding Remarks}
\label{sec:conclusion}

We have shown upper bounds on the error probabilities of the first
and the second kind,
based on a key operator inequality satisfied
by a density operator and its pinching.
The upper bounds are regarded as a noncommutative analogue
of the Hoeffding bound \cite{Hoeffding},
which is the optimal bound in the classical hypothesis testing,
and the upper bounds provide a simple proof of the direct part of
the quantum Stein's lemma.
Compared with \cite{Hayashi-Stein},
the proof is considerably simple and leads to the exponential convergence of
the error probability of the first kind.

The error exponents derived here do not seem to be natural,
since $\psibar(s)$ lacks symmetry between $\rho$ and $\sigma$
that the original hypothesis testing problem has.
One may introduce the following quantity
as a substitute for $\psibar(s)$
to keep the symmetry:
\begin{align*}
\max\Bigl\{
-\log\Tr\left[
\rho\,\sigma^{\frac{s}{2}}\rho^{-s}\sigma^{\frac{s}{2}}
\right],
-\log\Tr\left[
\sigma\,\rho^{\frac{s}{2}}\sigma^{-s}\rho^{\frac{s}{2}}
\right]
\Bigr\},
\end{align*}
and Theorem \ref{theorem:rate} still holds with the above quantity.
On the other hand, $\psi(s)$ and $\phi(a)$
defined in \eqref{def-psi} \eqref{def-phi}
seem to be probable functions for
the optimal rate function in quantum hypothesis testing,
and the following inequalities are expected to hold
\begin{align}
\limsup_{n\rightarrow\infty}\oneover{n}\log\alpha_n\left(S_n(a)\right)
&\le -\phi(a),
\label{yosou-alpha} \\
\limsup_{n\rightarrow\infty}\oneover{n}\log\beta_n\left(S_n(a)\right)
&\le -\bigl(\phi(a)+a\bigr),
\label{yosou-beta}
\end{align}
where
\begin{align}
S_n(a)\Def\proj{\rho_n - e^{na}\sigma_n>0}.
\end{align}
The question of whether the inequalities hold or not
seems to be difficult, however,
and is left open.

\appendices

\section{Definition of the Pinching}
\label{appendix:pinching}

In this appendix, we summarize the definition of the pinching
and some of its properties.
Given an operator $A\in\L(\H)$,
let $A=\sum_{i=1}^{v(A)} a_iE_i$ be its spectral decomposition,
where $v(A)$ is the number of eigenvalues of $A$
mutually different from others, and each $E_i$ is the projection
corresponding to an eigenvalue $a_i$.
The following map defined by using the PVM $E=\{E_i\}_{i=1}^{v(A)}$
is called the pinching:
\begin{align}
\E_{A}: B \in\L(\H) \longmapsto \E_{A}(B)
\Def\sum_{i=1}^{v(A)} E_i B E_i \in\L(\H).
\label{pinching1}
\end{align}
The operator $\E_{A}(B)$ is also called the pinching when no confusion
is likely to arise, and it is sometimes denoted as $\E_{E}(B)$.
It should be noted here that
$\E_{A}(B)$ commutes with $A$ and we have
\begin{align}
\Tr[BC] = \Tr\left[\E_{A}(B)C\right]
\label{pinching2}
\end{align}
for any operator $C\in\L(\H)$ commuting with $A$.

\section{Key Operator Inequality}
\label{appendix:key-inequality}

The following lemma was appeared in \cite{Hayashi-Stein},
and played an important role in this manuscript.
\vspace{2ex}
\begin{lemma}[Hayashi \cite{Hayashi-Stein}]
Given a PVM $M=\{M_i\}_{i=1}^{v(M)}$ on $\H$,
we have for $\forall\rho\in\S(\H)$
\begin{align}
\rho \le v(M)\E_M(\rho),
\end{align}
where $\E_M(\rho)$ is the pinching defined
in Appendix~\ref{appendix:pinching}.
\label{lemma:Hayashi}
\end{lemma}
\vspace{2ex}
We show another proof here for readers' convenience
by using the following operator convexity.
\vspace{2ex}
\begin{lemma}
Given a nonnegative operator $A\in\L(\H)$,
the following map is operator convex.
\begin{align}
f_A: X\in\L(\H) \longmapsto X^*AX\in\L(\H).
\end{align}
In other words, we have
\begin{align}
f_A(tX+(1-t)Y)\le tf_A(X)+(1-t)f_A(Y)
\end{align}
for $\forall X,Y\in\L(\H)$ and $0\le \forall t\le 1$.
\label{lemma:convexity}
\end{lemma}
\vspace{2ex}
\begin{proof}
The assertion is shown by a direct calculation as follows
\begin{align}
&tf_A(X)+(1-t)f_A(Y) - f_A(tX+(1-t)Y) \nonumber \\
&= tX^*AX + (1-t)Y^*AY
- [tX+(1-t)Y]^*A\,[tX+(1-t)Y] \nonumber \\
&= t(1-t) [X^*AX -X^*AY-Y^*AX+Y^*AY] \nonumber \\
&= t(1-t) (X-Y)^*A\,(X-Y) \nonumber \\
&\ge 0.
\end{align}
\end{proof}
\vspace{2ex}
\noindent
Now Lemma \ref{lemma:Hayashi} is verified
by using Lemma \ref{lemma:convexity} as follows
\begin{align}
\oneover{v(M)^2}\rho
&= \Biggl( \oneover{v(M)}\sum_{i=1}^{v(M)}M_i \Biggr)
 \,\rho\, \Biggl( \oneover{v(M)}\sum_{i=1}^{v(M)}M_i \Biggr) \nonumber \\
&\le \oneover{v(M)} \sum_{i=1}^{v(M)} M_i \rho M_i \nonumber \\
&= \oneover{v(M)}\E_{M}(\rho).
\end{align}

\section*{Acknowledgment}

The authors are grateful to Prof.~Hiroshi~Nagaoka.
He encouraged them to show a simple proof of the direct part
of the quantum Stein's lemma,
pointing out that the proof leads to Hiai-Petz's theorem.

This research was partially supported by
the Ministry of Education, Culture, Sports, Science, and Technology
Grant-in-Aid for Encouragement of Young Scientists, 13750058, 2001.


\newpage

\vspace{4ex}

\end{document}